\documentclass[aps,pra,preprint,superscriptaddress,showpacs,showkeys]{revtex4}
\usepackage{amssymb,bm}
\usepackage{graphicx}
\usepackage{amsmath}
\usepackage{epstopdf}
\usepackage{float}
\allowdisplaybreaks

\begin{document}

\title{Double   bremsstrahlung  from high-energy electron in the atomic  field.}

\author{P.A. Krachkov}\email{peter_phys@mail.ru}
\affiliation{Budker Institute of Nuclear Physics, 630090 Novosibirsk, Russia}
\affiliation{Novosibirsk State University, 630090 Novosibirsk, Russia}
\author{R. N. Lee}\email{R.N.Lee@inp.nsk.su}
\affiliation{Budker Institute of Nuclear Physics, 630090 Novosibirsk, Russia}
\author{A. I. Milstein}\email{A.I.Milstein@inp.nsk.su}
\affiliation{Budker Institute of Nuclear Physics, 630090 Novosibirsk, Russia}

\date{\today}

\begin{abstract}
The differential cross section of double   bremsstrahlung  from  high-energy electron in the electric  field of heavy atoms is derived. The results are obtained with the exact account  of the atomic field by means of  the quasiclassical approximation to the wave functions and the Green's function in the external field. It is shown that the Coulomb corrections to the differential cross section (the difference between the exact result and the result obtained in the leading Born approximation) correspond to  small momentum transfers. The Coulomb corrections  to the differential cross section of double   bremsstrahlung are accumulated in the factor, which coincides with the corresponding factor in the differential cross section of single  bremsstrahlung. At small momentum transfer, this factor is very sensitive to the parameters of screening while the Coulomb corrections to the spectrum have the universal form.
\end{abstract}



\maketitle

\section{Introduction}
To search for New Physics in precision experiments, it is necessary to know with high accuracy the cross sections of the main background processes. In particular, it is necessary to know  the cross sections  of single high-energy bremsstrahlung and  particle-antiparticle photoproduction  in the electric field of a heavy nucleus or  atom. These processes play a dominant role when considering electromagnetic showers in detectors.  In many cases they also give significant part of the radiative corrections.
In the  Born approximation, the   cross sections of both processes are known for arbitrary  energies of particles \cite{BH1934, Racah1934} (see also  Ref.~\cite{BLP1982}). However, for large $Z$ the Coulomb corrections (i.e. the contribution of higher-order terms in the parameter  $\eta=Z\alpha$) to the cross section are  very important (here  $Z$ is the atomic charge number, $\alpha=e^2\approx 1/137$ is the fine-structure constant, $e$ is the electron charge, $\hbar=c=1$). Though   there are formal expressions for the  cross sections exact in $\eta$ and energies of particles \cite{Overbo1968}, their use for numerical computations  becomes very difficult at high energies \cite{SudSharma2006}.

Fortunately, at high energies of initial particles, the main contribution to the cross section comes from small angles of the final particle momenta   with respect to the incident direction. In this case  typical angular momenta are large ($l\sim E/\Delta\gg 1$, where $E$ is energy and  $\Delta$ is the momentum transfer). Therefore,  the quasiclassical approximation, which accounts for large angular momenta  contributions,  becomes applicable.  Using the quasiclassical wave functions and the quasiclassical Green's functions of the Dirac equation in the external field, one can drastically simplify calculations. The celebrated Furry-Sommerfeld-Maue wave functions  \cite{Fu, ZM} (see also Ref.~\cite{BLP1982}) is nothing else but the leading-order quasiclassical wave functions for the Coulomb field. The quasiclassical Green's function  have been derived in Ref.~\cite{MS1983} for the case of a pure Coulomb field, in Ref.~\cite{LM95A} for an arbitrary spherically symmetric field, in Ref.~\cite{LMS00} for any localized field, and in Ref.~\cite{DM2012} for combined strong laser and atomic fields.

For   pair photoproduction and single bremsstrahlung, the cross sections in the leading  quasiclassical approximation  have been obtained in Refs.~\cite{BM1954,DBM1954,OlsenMW1957,O1955,OM1959}. The first quasiclassical corrections to the spectra of both processes have been obtained in Refs. \cite{LMS2004,LMSS2005,DM10,DM12}.  Recently, the  first quasiclassical corrections to the fully differential cross sections were obtained in Ref.~\cite{LMS2012} for $e^+e^-$  pair photoproduction, in Ref.~\cite{DLMR2014} for $\mu^+\mu^-$ pair  photoproduction, and in Ref.~\cite{KM2015} for single bremsstrahlung  from high-energy electrons and muons in an atomic  field.  The account for the first  quasiclassical corrections allows one to determine quantitatively the charge asymmetry in these processes (the asymmetry of the cross sections with respect to the permutation of particle and antiparticle). This asymmetry is absent in the cross section calculated in the leading quasiclassical approximation.

Influence of  screening (the difference between the atomic field and the Coulomb field of a nucleus) on the Coulomb corrections to  $e^+e^-$  pair photoproduction cross section is small  for the differential cross section and for the total cross section as well \cite{DBM1954}, see Ref.~\cite{LMS2004}, where the effect of screening has been investigated  quantitatively.  However, screening is important  for the Born term. A role   of screening in single  bremsstrahlung  in the atomic field is different.  It is shown in Refs.~\cite{OlsenMW1957,LMSS2005} that the Coulomb corrections to the differential cross section are very susceptible to screening. However, the Coulomb corrections to the cross section integrated over the momentum of final charged particle (electron or muon)
are independent of screening in the leading approximation over a small parameter $1/mr_{scr}$ \cite{LMSS2005}, where $r_{scr}\sim Z^{-1/3}(m\alpha)^{-1}$ is a screening
radius and  $m$ is the electron mass.

Investigation of high-energy $e^+e^-$ photoproduction  accompanied by  bremsstrahlung and double   bremsstrahlung from electrons in the electric field of a heavy atom (i.e., the processes $\gamma_1Z\rightarrow e^+e^-\gamma_2Z$ and $e^{\pm}Z\rightarrow \gamma_1\gamma_2e^{\pm}Z$, respectively)  is even more complicated task.  The  process
 $\gamma_1Z\rightarrow e^+e^-\gamma_2Z$ is a significant part of the radiative corrections to  $e^+e^-$ photoproduction as well as a noticeable background to such processes as Delbr\"uck scattering \cite{Delb1998}. This process should be taken into account at the consideration of the electromagnetic showers in the matter. During a long time only a few papers, related  to this process, have been published \cite{Huld1967,Corbo1978}. In those papers  the Born approximation was used.  Very recently, using the quasiclassical approximation,  the cross section of the process
 $\gamma_1Z\rightarrow e^+e^-\gamma_2Z$ at high energies  was  derived exactly in the parameter $\eta$  \cite{KLM2014}. It was shown that, apart from the region of very small momentum transfer, account of the Coulomb corrections for heavy atoms drastically change the result.

As to the double  bremsstrahlung cross section from electron in an atomic  field, it has been investigated either at low electron energies \cite{Korol93,KMMS2002} or for any electron energies but in the Born approximation \cite{Smirnov77}. In the present paper we use the quasiclassical approximation to derive the exact in $\eta$ differential cross section of  double  bremsstrahlung  from high energy electron in an atomic  field. We take into account the effect of screening and show that the Coulomb corrections to the cross section are, in general, very sensitive to this effect. Moreover, the Coulomb corrections to the double  bremsstrahlung cross section are accumulated in the factor which coincides with the corresponding factor in the differential cross section of single  bremsstrahlung. This allows us to formulate a recipe for the calculation of the  multiple bremsstrahlung amplitudes.

\section{General discussion}\label{general}
The differential cross section of double  bremsstrahlung  in the electric  field of a heavy atom  reads~\cite{BLP1982}
\begin{equation}\label{eq:cs}
d\sigma=\frac{\alpha^2}{(2\pi)^6}\omega_1\omega_2q\varepsilon_qd\omega_1d\omega_2\,d\Omega_{\bm{k}_1}\,d\Omega_{\bm{k}_2}d\Omega_{\bm{q}}\,|M|^{2}\,,
\end{equation}
where $d\Omega_{\bm{k}_1}$, $d\Omega_{\bm{k}_2}$, and $d\Omega_{\bm{q}}$ are  the solid angles corresponding to the photon momentum $\bm k_1$, $\bm k_2$, and the final charged particle momentum $\bm q$,
$\varepsilon_{ q}=\varepsilon_{ p}-\omega_1-\omega_2$ is the final charge particle energy, $\varepsilon_{ p}=\sqrt{ \bm{p}^2+m^2}$,  $\varepsilon_{ q}=\sqrt{ \bm q^2+m^2}$. Below we assume that $\varepsilon_{ p}\gg m$ and $\varepsilon_{ q}\gg m$.   The matrix element $M$ reads
\begin{align}\label{M12}
M&=M^{(1)}+M^{(2)}\,,\nonumber\\
M^{(1)}&=-\iint d\bm r_1d\bm r_2\, e^{-i\bm k_1\cdot\bm r_1-i\bm k_2\cdot\bm r_2 }\,\bar u_{\bm q }^{(-)}(\bm r_2 )\hat{e}^*_2G(\bm r_2,\bm r_1|\varepsilon_p-\omega_1)
\hat{e}^*_1\,u _{\bm p}^{(+)}(\bm r_1 )\,,\nonumber\\
M^{(2)}&=M^{(1)}(\bm k_1\leftrightarrow \bm k_2\,,\,\omega_1\leftrightarrow \omega_2\,,\, \bm e_1 \leftrightarrow\bm e_2)\, ,
\end{align}
where $\hat{e}=\gamma^\nu e_\nu=-\bm\gamma\cdot\bm e$,  $\gamma^\nu$ are the Dirac matrices, $ u_{\bm p}^{(+)}(\bm r )$  and $u_{\bm q}^{(-)}(\bm r )$ are the solutions of the Dirac equation in the  atomic potential  $V(r)$, $\bm e_{1\,2}$ are the photon polarization vectors, and $G(\bm r_2,\bm r_1|\varepsilon)$ is the Green's function of the Dirac equation in the potential $V(r)$. The superscripts $(-)$ and $(+)$ remind us that the asymptotic forms of
 $ u_{\bm q}^{(-)}(\bm r )$ and $ u_{\bm p}^{(+)}(\bm r )$ at large $\bm r$ contain, in addition to the plane wave, the spherical convergent and divergent waves, respectively. It is convenient to write
the contribution $M^{(1)}$ in  Eq.~(\ref{M12}) in terms of the Green's function $D(\bm r_2,\,\bm r_1|\varepsilon)$ of the  ``squared'' Dirac equation,
\begin{equation}\label{GD}
G(\bm r_2,\,\bm r_1|\varepsilon)=(\hat{\cal P}+m)D(\bm r_2,\,\bm r_1|\varepsilon),\quad D(\bm r_2,\,\bm r_1|\varepsilon)=\langle \bm r_2|\frac{1}{\hat{\cal P}^2-m^2+i0}| \bm r_1\rangle\,,
\end{equation}
where $\hat{\cal P}=\gamma^\nu{\cal P}_\nu$, ${\cal P}_\nu=\big(\varepsilon-V(r),i\bm\nabla\big)$. Substituting Eq.~\eqref{GD} in Eq.~\eqref{M12}, performing integration by parts and using the Dirac equation, we obtain
\begin{align}\label{M1D}
M^{(1)}&=-\iint d\bm r_1d\bm r_2\,e^{-i\bm k_1\cdot\bm r_1-i\bm k_2\cdot\bm r_2 }\,\bar u_{\bm q }^{(-)}(\bm r_2 )\hat{e}^*_2D(\bm r_2,\bm r_1|\varepsilon_p-\omega_1)\nonumber\\
&\times[2i\bm {e}^*_1\cdot\bm\nabla+\hat{e}^*_1\hat{k}_1]\,u _{\bm p}^{(+)}(\bm r_1 )\,.
\end{align}
The Green's function  $D(\bm r_2,\,\bm r_1|\varepsilon)$ and the wave functions $ u_{\bm p}^{(+)}(\bm r )$  and $u_{\bm q}^{(-)}(\bm r )$ have the form \cite{DLMR2014,KM2015}
\begin{align}\label{wfD1}
&D(\bm r_2,\,\bm r_1|\varepsilon)=d_0(\bm r_2,\bm r_1)+\bm\alpha\cdot\bm d_1(\bm r_2,\bm r_1)+\bm\Sigma\cdot\bm d_2(\bm r_2,\bm r_1)\,,\nonumber\\
&\bar u_{\bm q }^{(-)}(\bm r )=\bar u_{\bm q }[f_0(\bm r,\bm q)-\bm\alpha\cdot\bm f_1(\bm r,\bm q)-\bm\Sigma\cdot\bm f_2(\bm r,\bm q)]\,,\nonumber\\
&u _{\bm p}^{(+)}(\bm r )=[g_0(\bm r,\bm p)-\bm\alpha\cdot\bm g_1(\bm r,\bm p)-\bm\Sigma\cdot\bm g_2(\bm r,\bm p)]u _{\bm p}\,,\nonumber\\
& u_{\bm p}=\sqrt{\frac{\varepsilon_p+m}{2\varepsilon_p}}
\begin{pmatrix}
\phi\\
\dfrac{{\bm \sigma}\cdot {\bm
p}}{\varepsilon_p+m}\phi
\end{pmatrix}\,,\quad
u_{\bm q}=\sqrt{\frac{\varepsilon_q+m}{2\varepsilon_q}}
 \begin{pmatrix}
\chi\\
\dfrac{{\bm \sigma}\cdot {\bm q}}{\varepsilon_q+m}\chi
\end{pmatrix}\,,
\end{align}
where $\phi$ and  $\chi$  are spinors, $\bm\alpha=\gamma^0\bm\gamma$, $\bm\Sigma=\gamma^0\gamma^5\bm\gamma$, $\gamma^5=-i\gamma^0\gamma^1\gamma^2\gamma^3$, and $\bm\sigma$ are the Pauli matrices.
The coefficients $d_0$, $\bm d_{1}$, $f_0$, $\bm f_{1}$,  $g_0$ and $\bm g_{1}$ in the leading quasiclassical approximation, as well as the first quasiclassical corrections to  $d_0$,  $f_0$ and  $g_0$, were derived in Ref.~\cite{LMS00} for arbitrary atomic potential $V(r)$. The first quasiclassical corrections to  $\bm d_{1}$, $\bm f_{1}$ and $\bm g_{1}$, together with the leading quasiclassical terms of  $\bm d_{2}$, $\bm f_{2}$ and $\bm g_{2}$, were derived in Ref.~\cite{KM2015}. We perform calculation of the double   bremsstrahlung cross section in the leading  quasiclassical approximation. In this case it is sufficient to take into account the terms $d_0$, $\bm d_{1}$, $f_0$, $\bm f_{1}$,  $g_0$ and $\bm g_{1}$ in the leading quasiclassical approximation and neglect the contributions of
 $\bm d_{2}$, $\bm f_{2}$ and $\bm g_{2}$ \cite{LMS2012,DLMR2014,KM2015}.  Within this accuracy we have for  $d_0$ and $\bm d_{1}$
\begin{align}\label{dqc}
&d_0(\bm r_2,\bm r_1)=
\frac{ie^{i\kappa r}}{4\pi^2r} \int d\bm Q \exp\left[iQ^2-i
r\int_0^1dx V(\bm R) \right]\, ,\nonumber\\
&\bm d_1(\bm r_2,\bm r_1)=-\frac{i}{2\varepsilon}(\bm \nabla_1+\bm \nabla_2)d_0(\bm r_2,\bm r_1)\, ,\nonumber\\
&\bm r =\bm r_2-\bm r_1\,,\quad \bm R= \bm r_1+x\bm r+\bm Q \sqrt{\frac{2x(1-x)r}{\kappa}}\,,\quad \kappa=\sqrt{\varepsilon^2-m^2}\,,
\end{align}
where $\bm Q$ is a two-dimensional vector perpendicular to the vector $\bm r_2-\bm r_1$. The terms  $f_0$ and  $\bm f_{1}$ are
\begin{align}\label{fgqc}
& f_0(\bm r,\bm q)=-\frac{i}{\pi}e^{-i\bm q\cdot\bm r}\int d\bm Q \exp\left[iQ^2-i
\int_0^\infty dx V(\bm r_q) \right]\,,\nonumber\\
&\bm f_1(\bm r,\bm q)=\frac{1}{2\varepsilon_q}(i\bm\nabla-\bm q)f_0(\bm r,\bm q)\,,\nonumber\\
&\bm r_q= \bm r+x\bm n_{ q}+\bm Q \sqrt{\frac{2x}{\varepsilon_q}}\,, \quad \bm Q\cdot\bm n_{ q}=0\,,\quad \bm n_{ q}=\bm q/q\,.
\end{align}
The expressions for  $g_0$ and $\bm g_{1}$ follow from the relations
\begin{align}\label{gvf}
& g_0(\bm r,\bm p)=f_0(\bm r,-\bm p)\,,\quad \bm g_1(\bm r,\bm p)=\bm f_1(\bm r,-\bm p)\,.
\end{align}
It is convenient to calculate the matrix element for definite helicities of the particles. Let $\mu_p$, $\mu_q$, $\lambda_1$, and $\lambda_2$ be the signs of the helicities of  initial electron, final  electron, and radiated photons, respectively.   We fix the coordinate system so that $\bm\nu\equiv\bm n_{p}=\bm p/p$  is directed along $z$-axis and $\bm q$ lies in the $xz$ plane with $q_{x}>0$. Denoting helicities by the subscripts, we have
\begin{align}\label{spinors}
\phi&=\frac{1+\mu_p\bm \sigma\cdot\bm n_p}{4}
\begin{pmatrix}1+\mu_p\\1-\mu_p\end{pmatrix}\,,
\nonumber\\
\chi&=\frac{1+\mu_q\bm \sigma\cdot\bm n_q}{4\cos(\theta_{q}/2)}
\begin{pmatrix}1+\mu_q\\1-\mu_q\end{pmatrix}
\approx \frac14\left(1+\frac{\theta_{q}^2}{8}\right)\left(1+\mu_q\bm \sigma\cdot\bm n_q\right)\begin{pmatrix}1+\mu_q\\1-\mu_q\end{pmatrix}\,,
\nonumber\\
\bm e_{1}&=  \bm s_{\lambda_1}-(\bm s_{\lambda_1}\cdot\bm\theta_{k_1})\bm\nu\,,\quad \bm e_{2}=  \bm s_{\lambda_2}-(\bm s_{\lambda_2}\cdot\bm\theta_{k_2})\bm\nu\,,\nonumber\\
\bm s_{\lambda}&=\frac{1}{\sqrt{2}}(\bm e_x+i\lambda\bm e_y)\,,
\end{align}
where  $\bm\theta_q=\bm q_\perp/q$, $\bm\theta_{k_1}=\bm k_{1\perp}/\omega_1$, and $\bm\theta_{k_2}=\bm k_{2\perp}/\omega_2$,
the notation  $\bm X_\perp=\bm X-(\bm\nu\cdot\bm X)\bm\nu$ for any vector $\bm X$ is used. Below we assume that  $\theta_{q}\ll 1$,  $\theta_{k_1}\ll 1$, and  $\theta_{k_2}\ll 1$. The unit vectors $\bm e_x$ and $\bm e_y$ are directed along $\bm q_{\perp}$  and $\bm p \times \bm q$. In the expressions for $\bm e_{1}$ and $\bm e_{2}$ in \eqref{spinors}, the terms of the order $O(\theta_{k_1}^2)$ and  $O(\theta_{k_2}^2)$ are omitted. For  the matrix  ${\cal F}=u_{{\bm p}\,\mu_p}\bar{u}_{{\bm q}\,\mu_q}$ we have \cite{KM2015}
\begin{eqnarray}\label{calf}
&&{\cal F}=\frac{1}{8}(a_{\mu_p\mu_q}+\bm\Sigma\cdot \bm b_{\mu_p\mu_q})[\gamma^0(1+PQ)+\gamma^0\gamma^5(P+Q)+(1-PQ)-\gamma^5(P-Q)],\nonumber\\
&&P=\frac{\mu_pp}{\varepsilon_p+m}\,,\quad Q=\frac{\mu_qq}{\varepsilon_q+m}\,.
\end{eqnarray}
Here $a_{\mu_p\mu_q}$ and $\bm b_{\mu_p\mu_q}$ are
\begin{eqnarray}\label{ab}
&&a_{\mu\mu}=1-\frac{\theta_{q}^2}{8}\,,\quad  a_{\mu\bar\mu}=-\frac{\mu}{\sqrt{2}}\bm s_\mu\cdot\bm\theta_{q}\,,\nonumber\\
&&\bm b_{\mu\mu}=\mu\left(1-\frac{\theta_{q}^2}{8}\right)\bm\nu+\frac{\mu}{2}\bm\theta_{q}-\frac{i}{2}[\bm\theta_{q}\times\bm\nu]\,,\nonumber\\
&&\bm b_{\mu\bar\mu}=\sqrt{2}\bm s_\mu-\frac{1}{\sqrt{2}}(\bm s_\mu\cdot\bm\theta_{q})\bm\nu\,,\quad \bm s_{\mu}=\frac{1}{\sqrt{2}}(\bm e_x+i\mu\bm e_y)\,,
\end{eqnarray}
where  $\bar\mu=-\mu$. The matrix element $M$, Eq.~\eqref{M1D}, can be written as follows
\begin{align}\label{MF}
M^{(1)}&=-\iint d\bm r_1d\bm r_2\,e^{-i\bm k_1\cdot\bm r_1-i\bm k_2\cdot\bm r_2 }\,\mbox{Tr}\,[f_0\hat{e}^*_2d_0\Theta\, g_0-\bm\alpha\cdot\bm f_1\hat{e}^*_2d_0\Theta\, g_0\nonumber\\
&+f_0\hat{e}^*_2\bm\alpha\cdot\bm d_1\Theta\, g_0-f_0\hat{e}^*_2d_0\Theta\,\bm\alpha\cdot\bm g_1]{\cal F}\,,\nonumber\\
&\Theta=2i\bm {e}^*_1\cdot\bm\nabla+\hat{e}^*_1\hat{k}_1\,.
\end{align}
Here the functions $d_0$ and $\bm d_1$ are calculated at $\varepsilon=\varepsilon_p-\omega_1$.
Note that only the terms with  $(P+Q)$ and $(1+PQ)$ in $\cal F$, Eq.~\eqref{calf}, contribute to the matrix element (\ref{MF}) due to the trace over $\gamma$-matrices. Below we calculate the matrix element $M$ for the atomic potential $V(r)$, which includes the effect of screening.

\section{Matrix element and cross section}\label{matrixelement}

The calculation of the matrix element  \eqref{M12} is performed in the same way as  in Ref.\cite{LMSS2005}. Some details of this calculation are given in Appendix.
The final result is:
\begin{align}\label{M12F}
&M_{\mu_p\mu_q\lambda_1\lambda_2}=-\bm A(\bm\Delta)\cdot\Big [\bm T_{\mu_p\mu_q\lambda_1\lambda_2}(\bm k_1,\bm k_2)
+\bm T_{\mu_p\mu_q\lambda_2\lambda_1}(\bm k_2,\bm k_1)\Big]\,,\nonumber\\
&\bm A(\bm\Delta) =-i\int d\bm r \exp\left[-i\bm\Delta\cdot\bm r-i\chi(\rho)\right]\bm\nabla_\perp V(r)\,,\quad \chi(\rho)=\int_{-\infty}^\infty V(\sqrt{z^2+\rho^2})dz\,,  \nonumber\\
&\bm T_{++++}(\bm k_1,\bm k_2)=p\left[(\bm e^*\cdot\bm\theta_{k_1})(\bm e^*\cdot\bm\theta_{k_2 q})\bm j_0+N_1(\bm e^*\cdot\bm\theta_{k_1})\bm e^*+
N_3(\bm e^*\cdot\bm\theta_{ k_2 q})\bm e^*\right]\,,\nonumber\\
&\bm T_{+++-}(\bm k_1,\bm k_2)=\left[p(\bm e^*\cdot\bm\theta_{k_1})(\bm e\cdot\bm\theta_{k_2})-\frac{m^2\omega_1}{2pq}\right]\bm j_0+p(N_2+N_3)(\bm e^*\cdot\bm\theta_{k_1})\bm e\nonumber\\
&+N_3(\bm e,\, p\bm\theta_{k_2}-\bm\Delta_\perp)\bm e^*\,,\nonumber\\
&\bm T_{++-+}(\bm k_1,\bm k_2)=\varkappa\left[(\bm e\cdot\bm\theta_{ k_1})(\bm e^*\cdot\bm\theta_{k_2q})\bm j_0+N_1(\bm e\cdot\bm\theta_{ k_1})\bm e^*+
N_3(\bm e^*\cdot\bm\theta_{k_2q})\bm e\right]\,,\nonumber\\
&\bm T_{++--}(\bm k_1,\bm k_2)=q\left[(\bm e\cdot\bm\theta_{k_1})(\bm e\cdot\bm\theta_{k_2 q})\bm j_0+N_1(\bm e\cdot\bm\theta_{k_1})\bm e+
N_3(\bm e\cdot\bm\theta_{ k_2 q})\bm e\right]\,,\nonumber\\
&\bm T_{+-++}(\bm k_1,\bm k_2)=-\frac{m(\omega_1+\omega_2)}{\sqrt{2}q}\left[(\bm e^*\cdot\bm\theta_{k_1})\bm j_0+N_3\bm e^*\right]
-\frac{m \omega_1 }{\sqrt{2}p}\left[(\bm e^*\cdot\bm\theta_{k_2k_1})\bm j_0+N_2\bm e^*\right]\,,\nonumber\\
&\bm T_{+-+-}(\bm k_1,\bm k_2)=-\frac{m\omega_1}{\sqrt{2}p}\left[(\bm e\cdot\bm\theta_{k_2q})\bm j_0+N_1\bm e\right]\,,\nonumber\\
&\bm T_{+--+}(\bm k_1,\bm k_2)=-\frac{m\omega_2}{\sqrt{2}q}\left[(\bm e\cdot\bm\theta_{k_1})\bm j_0+N_3\bm e\right]\,,\nonumber\\
&\bm T_{+---}(\bm k_1,\bm k_2)=0\,, \nonumber\\
& \bm T_{{\bar\mu}_p{\bar\mu}_q{\bar\lambda}_1{\bar\lambda}_2}(\bm k_1,\bm k_2)=
{\bar\mu}_p{\bar\mu}_q\bm T_{\mu_p\mu_q\lambda_1\lambda_2}(\bm k_1,\bm k_2)\,|_{\bm e\leftrightarrow \bm e^*}\,,\quad{\bar\mu}=-\mu\,,\quad  {\bar\lambda}=-\lambda\,.
\end{align}
Here we use the following notation
\begin{align}\label{notation}
&\bm e=\frac{1}{\sqrt{2}}(\bm e_x+i\bm e_y)\,,\quad \bm\theta_{k_2q}=\bm\theta_{k_2}-\bm\theta_q\,,\quad \bm\theta_{k_2k_1}=\bm\theta_{k_2}-\bm\theta_{k_1}\,,\nonumber\\
&\varkappa=p-\omega_1\,,\quad\bm\Delta=\bm q+\bm k_1+\bm k_2-\bm p\,,\quad \bm\Delta_\perp=q\bm\theta_q+\omega_1\bm\theta_{k_1}+\omega_2\bm\theta_{k_2}\,, \nonumber\\
&\bm j_0=\frac{4}{a_1a_2a_3a_4}\{a_3[(p+q)\bm\Delta_\perp-2pq\bm\theta_{q}]+a_1\omega_2(\bm\Delta_\perp+2q\bm\theta_{k_2q})\}\,,\nonumber\\
&N_1=\frac{4}{a_2a_3}\,,\quad N_2=\frac{4}{a_3a_4}\,,\quad N_3=\frac{4}{a_1a_4}\,,\nonumber\\
&a_1=-\frac{\omega_1}{\varkappa}[(\bm\Delta_\perp-p\bm\theta_{k_1})^2+m^2]-\frac{p\omega_2}{q\varkappa}[q^2\bm\theta_{k_2q}^2+m^2]\,,\nonumber\\
&a_2=\frac{\omega_2}{\varkappa}[(\bm\Delta_\perp+q\bm\theta_{k_2q})^2+m^2]+\frac{q\omega_1}{p\varkappa}[p^2\bm\theta_{k_1}^2+m^2]\,,\nonumber\\
&a_3=\frac{\omega_1}{p}[p^2\bm\theta_{k_1}^2+m^2]\,,\quad a_4=\frac{\omega_2}{q}[q^2\bm\theta_{k_2q}^2+m^2]\,.
\end{align}
Within our accuracy, one can replace $p$ and $q$ in  Eqs.~\eqref{M12F} and \eqref{notation} by $\varepsilon_p$ and $\varepsilon_q$. The vector $\bm A(\bm\Delta)$ is obviously parallel to the vector $\bm\Delta_\perp$,
\begin{align}\label{AA0}
\bm A(\bm\Delta)=A_0(\bm\Delta)\,\bm\Delta_\perp\,,\quad A_0(\bm\Delta)=-\frac{i}{\Delta_\perp^2}\int d\bm r
\exp\left[-i\bm\Delta\cdot\bm r-i\chi(\rho)\right]\bm\Delta_\perp\cdot\bm\nabla_\perp V(r)\,,
\end{align}
so that we can write the amplitude $M_{\mu_p\mu_q\lambda_1\lambda_2}$ as
\begin{align}\label{MFA0}
&M_{\mu_p\mu_q\lambda_1\lambda_2}=-A_0(\bm\Delta)\,{\cal T}_{\mu_p\mu_q\lambda_1\lambda_2}\,,\nonumber\\
&{\cal T}_{\mu_p\mu_q\lambda_1\lambda_2}=\bm\Delta_\perp\cdot\Big [\bm T_{\mu_p\mu_q\lambda_1\lambda_2}(\bm k_1,\bm k_2)
+\bm T_{\mu_p\mu_q\lambda_2\lambda_1}(\bm k_2,\bm k_1)\Big]\,.
\end{align}
The amplitude $M_{\mu_p\mu_q\lambda_1\lambda_2}$ is exact in the potential $V(r)$. Whole  dependence of this amplitude on the potential $V(r)$ is contained in the factor $A_0(\bm\Delta)$.
In the Born approximation we have
\begin{equation}\label{AB}
A_{0}^B(\bm\Delta)=V_F(\Delta^2)=-\frac{4\pi\eta{ F}(\Delta^2)}{\Delta^2}\,,
\end{equation}
 where $V_F(\Delta^2)$ it the Fourier transformation of the potential $V(r)$, and  ${F}(\Delta^2)$ is the atomic form factor, which differs essentially  from unity at $\Delta\lesssim 1/r_{scr}$.
Thus, the Born amplitude reads
\begin{align}\label{MB}
&M_{\mu_p\mu_q\lambda_1\lambda_2}^B=-V_F(\Delta^2)\,{\cal T}_{\mu_p\mu_q\lambda_1\lambda_2}\,,
\end{align}
where ${\cal T}_{\mu_p\mu_q\lambda_1\lambda_2}$ coincides with that in Eq.~\eqref{MFA0}.

If $\Delta_\perp\gg \mbox{max}(r_{scr}^{-1},|\Delta_{\parallel}|)$ then we can neglect the effect of screening, replace  $V(r)$ by the Coulomb potential $V_c(r)=-\eta/r$,
and neglect also $\Delta_\parallel=\bm \nu \cdot \bm \Delta$.
Within our precision
\begin{align}\label{delpar}
\Delta_{\parallel}=-\frac{1}{2}\left[q\theta_q^2+\omega_1\theta_{k_1}^2+\omega_2\theta_{k_2}^2+\frac{m^2(\omega_1+\omega_2)}{pq}\right]\,.
\end{align}
A simple calculation gives for the factor $A_{0}(\bm\Delta)$:
 \begin{equation}\label{Aeik}
A_0(\bm\Delta)=-\frac{4\pi\eta(L\Delta)^{2i\eta}}{\Delta^2}\frac{\Gamma(1-i\eta)}{\Gamma(1+i\eta)}\,,
\end{equation}
where $\Gamma(x)$ is the Euler $\Gamma$ function and $L\sim \mbox{min}(|\Delta_{\parallel}|^{-1},\, r_{scr})$. Note that the factor $(L\Delta)^{2i\eta}$ is irrelevant because it disappears in $|M|^2$.
Thus, in the  region  $\Delta_\perp\gg \mbox{max}(r_{scr}^{-1},|\Delta_{\parallel}|)$, we have   $|A_0(\bm\Delta)|=|A_0^B(\bm\Delta)|$.

Let us represent the cross section $d\sigma$ \eqref{eq:cs} as a sum of the Born term and the Coulomb corrections:
\begin{align}\label{sigBC}
&d\sigma_{\mu_p\mu_q\lambda_1\lambda_2}=d\sigma^B_{\mu_p\mu_q\lambda_1\lambda_2}+d\sigma^C_{\mu_p\mu_q\lambda_1\lambda_2}\,,\nonumber\\
&d\sigma^B_{\mu_p\mu_q\lambda_1\lambda_2}=\frac{\alpha^2}{(2\pi)^6}\omega_1\omega_2\,d\omega_1d\omega_2\,d\bm\theta_{k_1}\,d\bm\theta_{k_2}d\bm\Delta_\perp\,|A_0^B(\bm\Delta)|^{2}
|{\cal T}_{\mu_p\mu_q\lambda_1\lambda_2}|^2   \,,\nonumber\\
&d\sigma^C_{\mu_p\mu_q\lambda_1\lambda_2}=\frac{\alpha^2}{(2\pi)^6}\omega_1\omega_2\,d\omega_1d\omega_2\,d\bm\theta_{k_1}\,d\bm\theta_{k_2}d\bm\Delta_\perp\,
R(\bm\Delta)\,|{\cal T}_{\mu_p\mu_q\lambda_1\lambda_2}|^2   \,,\nonumber\\
&R(\bm\Delta)=|A_0(\bm\Delta)|^{2}-|A_0^B(\bm\Delta)|^{2}\,,
\end{align}
where we pass from the integration over $d\Omega_{\bm{q}}$ to the integration over $d\bm\Delta_\perp$. It is seen from Eqs.~\eqref{AB} and \eqref{Aeik} that only the region of small $\Delta_\perp$,
$\Delta_\perp\sim \mbox{max}(r_{scr}^{-1},|\Delta_{\parallel}|)\ll m$,  gives the contribution to $d\sigma^C$. The term $\bm A(\bm\Delta)$ coincides with the corresponding term in the single bremsstrahlung cross section \cite{LMSS2005}. As shown in Ref.~\cite{LMSS2005}, the function $R(\bm\Delta)$ is very sensitive to the shape of the atomic potential at $r\sim r_{scr}$, while the integral,
\begin{align}\label{intR}
&\int d\bm\Delta_\perp\,\bm\Delta_\perp^2R(\bm\Delta)= -32\pi^3\eta^2f(\eta)\,,\nonumber\\
&f(\eta)=\mbox{Re}\,\psi(1+i\eta)-\psi(1)\,,
\end{align}
is independent of this shape; $\psi(x)=d\ln\Gamma(x)/dx$. Therefore, the Coulomb corrections integrated over $d\bm\Delta_\perp$ have the form
\begin{align}\label{sigCC}
&d\sigma^C_{\mu_p\mu_q\lambda_1\lambda_2}=-\frac{\alpha^2\eta^2f(\eta)}{4\pi^3}\omega_1\omega_2\,d\omega_1d\omega_2\,d\bm\theta_{k_1}\,d\bm\theta_{k_2}
\,|\bm T_{\mu_p\mu_q\lambda_1\lambda_2}^{(0)}(\bm k_1,\bm k_2)+\bm T_{\mu_p\mu_q\lambda_2\lambda_1}^{(0)}(\bm k_2,\bm k_1)|^2   \,,
\end{align}
where the function $\bm T_{\mu_p\mu_q\lambda_1\lambda_2}^{(0)}(\bm k_1,\bm k_2)$ is   $\bm T_{\mu_p\mu_q\lambda_1\lambda_2}(\bm k_1,\bm k_2)$, Eq.~\eqref{M12F}, taken at $\bm\Delta_\perp=0$, i.e., at
$\bm\theta_q=-(\omega_1\bm\theta_{k_1}+\omega_2\bm\theta_{k_2})/q$. The main contribution to the Born cross section integrated over $d\bm\Delta_\perp$ is given by the region $m\gg \Delta_\perp\gg m\beta$ of small $\Delta_\perp$, where
\begin{equation}\label{beta}
\beta=\mbox{max}\left\{\frac{1}{mr_{scr}}\,,\, \frac{|\Delta_{\parallel}|}m\right\}\,.
\end{equation}
Assuming that $\ln(1/\beta)\gg 1$,   we have within logarithmic accuracy:
\begin{align}\label{sigB}
&d\sigma^B_{\mu_p\mu_q\lambda_1\lambda_2}=\frac{\alpha^2\eta^2}{4\pi^3}\omega_1\omega_2\,d\omega_1d\omega_2\,d\bm\theta_{k_1}\,d\bm\theta_{k_2}\,\ln\frac{1}{\beta}\nonumber\\
&\times|\bm T_{\mu_p\mu_q\lambda_1\lambda_2}^{(0)}(\bm k_1,\bm k_2)+\bm T_{\mu_p\mu_q\lambda_2\lambda_1}^{(0)}(\bm k_2,\bm k_1)|^2   \,.
\end{align}

\begin{figure}[h]
\centering
\includegraphics[width=0.6\linewidth]{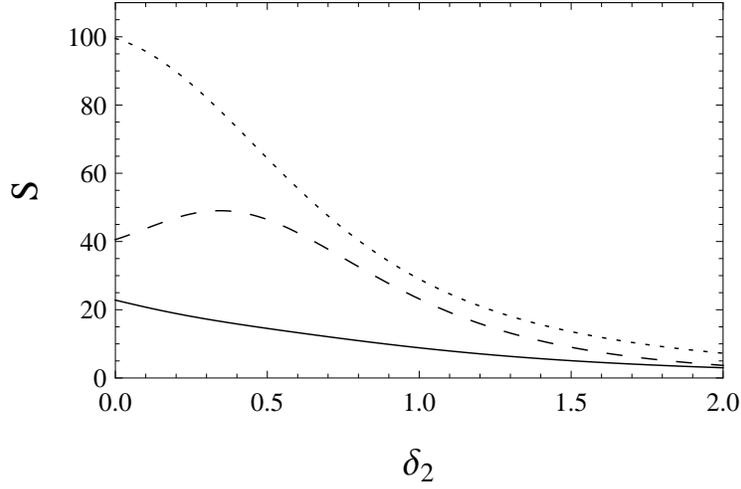}
\setlength{\unitlength}{0.7\linewidth}
\caption{The quantity  $S$ \eqref{S} as a function of $\delta_2=p\theta_{k_2}/m$ at $\omega_1/\varepsilon_p=0.2$, $\omega_2/\varepsilon_p=0.4$, $\phi=0$,   $\delta_1=p\theta_{k_1}/m=0.2$ (dashed curve),
$\delta_1=1$ (dotted curve), and $\delta_1=2$  (solid curve).}
\label{figtheta}
\end{figure}

In order to demonstrate the angular dependence of the Coulomb corrections, we introduce the dimensionless quantity $S$,
\begin{equation}\label{S}
S=\frac{m^6}{2}\sum_{\mu_p\mu_q\lambda_1\lambda_2}|\bm T_{\mu_p\mu_q\lambda_1\lambda_2}^{(0)}(\bm k_1,\bm k_2)+\bm T_{\mu_p\mu_q\lambda_2\lambda_1}^{(0)}(\bm k_2,\bm k_1)|^2\,,
\end{equation}
and show in Fig.~\ref{figtheta} the dependence of $S$ on $\delta_2=p\theta_{k_2}/m$ at fixed $\delta_1=p\theta_{k_1}/m$, $\omega_1/\varepsilon_p$, $\omega_2/\varepsilon_p$, and the azimuth angle $\phi$ between vectors $\bm \theta_{k_1}$ and $\bm \theta_{k_2}$.

\begin{figure}[h]
\centering
\includegraphics[width=0.6\linewidth]{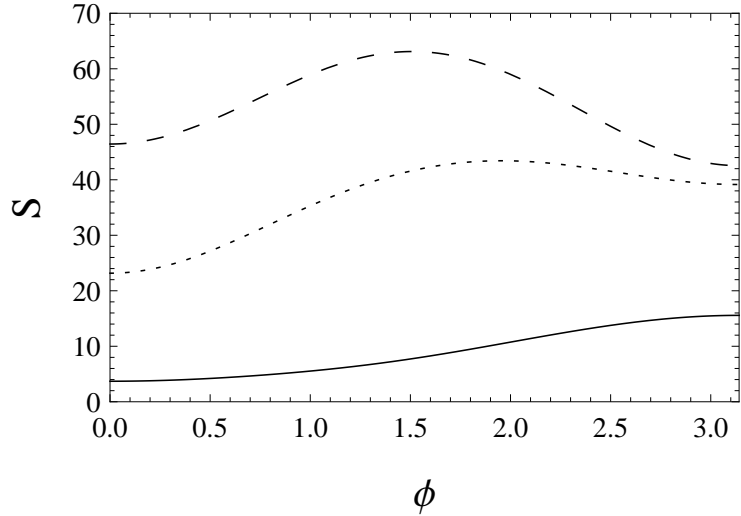}
\setlength{\unitlength}{0.7\linewidth}
\caption{The quantity  $S$ \eqref{S} as a function of the azimuth angle $\phi$ between vectors $\bm \theta_{k_1}$ and $\bm \theta_{k_2}$ at
$\omega_1/\varepsilon_p=0.2$, $\omega_2/\varepsilon_p=0.4$,  $\delta_1=0.2$,
$\delta_2=0.5$ (dashed curve), $\delta_2=1$(dotted curve), and $\delta_2=2$  (solid curve).}
\label{figphi}
\end{figure}

In Fig.~\ref{figphi} the quantity $S$ is shown as a function of $\phi$ at fixed $\delta_1=p\theta_{k_1}/m$, $\delta_2=p\theta_{k_2}/m$, $\omega_1/\varepsilon_p$,  and $\omega_2/\varepsilon_p$. Note that $S$ is invariant under the replacement $\phi\rightarrow -\phi$.

\begin{figure}[h]
\centering
\includegraphics[width=0.6\linewidth]{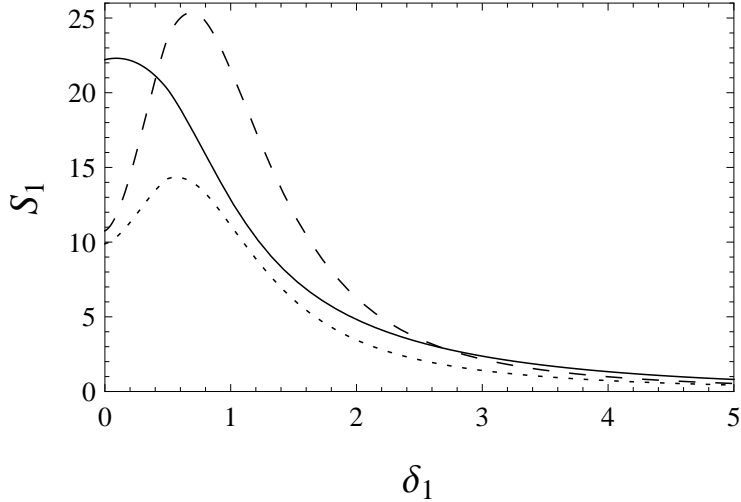}
\setlength{\unitlength}{0.7\linewidth}
\caption{The quantity  $S_1$ \eqref{S1} as a function of  $\delta_1 $ at $\omega_1/\varepsilon_p=\Omega x$ and $\omega_2/\varepsilon_p=\Omega (1-x)$, where $\Omega=0.4$, $x=0.3$ (dashed curve), $x=0.5$
(dotted curve), and  $x=0.7$ (solid curve).}
\label{fig1theta}
\end{figure}

It is seen from Figs.~\ref{figtheta} and  \ref{figphi} that $S$ has a smooth angular dependence. In Fig.~\ref{fig1theta} we show the dependence of the quantity $S_1$ on $\delta_1$ at fixed
 $\omega_1/\varepsilon_p$ and  $\omega_2/\varepsilon_p$, where
\begin{equation}\label{S1}
S_1=\frac{p^2}{16 \pi^2m^2}\int S\, d\bm\theta_{k_2}\,.
\end{equation}

It is seen that the main contribution to the cross section is given by the region $\delta_{1}\sim 1$.

\begin{figure}[h]
\centering
\includegraphics[width=0.6\linewidth]{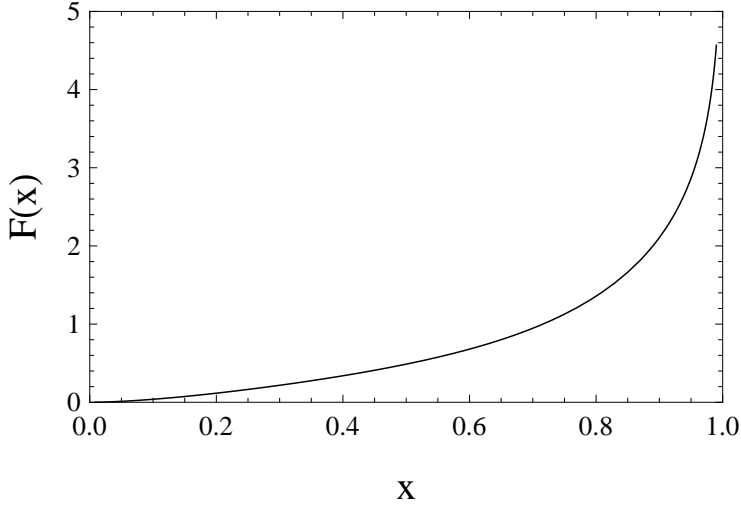}
\setlength{\unitlength}{0.7\linewidth}
\caption{Dependence  of $F(x)$, Eq.~\eqref{spectrumsoft},  on $x=\omega_1/\varepsilon_p$.}
\label{F}
\end{figure}

Let us discuss now the Coulomb corrections to the cross section integrated over $\bm\theta_{k_1}$ and  $\bm\theta_{k_2}$ ( the spectrum), averaged over the polarization of the initial electron polarization, and summed over polarizations of the final particles. We write it  as
\begin{align}\label{spectrumCC}
&d\sigma^C=-\frac{8\alpha^2\eta^2f(\eta)d\omega_1d\omega_2}{\pi m^2\omega_1\omega_2}G(\omega_1/\varepsilon_p,\, \omega_2/\varepsilon_p )\,,
\end{align}
where the function $f(\eta)$ is qiven in Eq.~\eqref{intR}. For $\omega_2\ll \omega_1,\,\varepsilon_q$, a  simple calculation gives the result, which corresponds to the soft-photon-emission  approximation \cite{BLP1982}:
\begin{align}\label{spectrumsoft}
&F(x)=G(x,0)=\int_0^\infty \frac{dy}{(1+y)^2}\left[1+(1-x)^2-\frac{4y(1-x)}{(1+y)^2}\right]\Phi(x,y)\,,\nonumber\\
&\Phi(x,y)=\frac{t}{\sqrt{t^2-1}}\ln(t+\sqrt{t^2-1})-1\,,\quad t=1+\frac{x^2(1+y)}{2(1-x)}\,.
\end{align}
The function $F(x)$ is shown in Fig.~\ref{F}.
The asymtotic behavior  of the function $F(x)$ is
\begin{align}
&F(x)\approx \frac{4}{3}x^2\ln\frac{1}{x}\,\quad \mbox{at}\quad x\ll 1\,,\nonumber\\
&F(x)\approx \ln\frac{1}{1-x}\,\quad \mbox{at}\quad 1-x\ll 1\,.
\end{align}

\begin{figure}[h]
\centering
\includegraphics[width=0.6\linewidth]{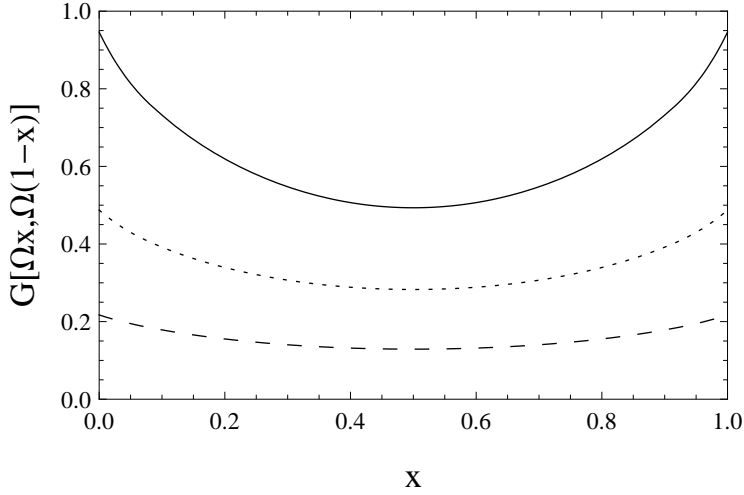}
\setlength{\unitlength}{0.7\linewidth}
\caption{Dependence of $G[\Omega x,\Omega (1-x)]$, Eq.~\eqref{spectrumsoft}, on $x$ at $\Omega=0.3$ (dashed curve), $\Omega=0.5$ (dotted curve), and $\Omega=0.7$  (solid curve).
Here $\Omega=(\omega_1+\omega_2)/\varepsilon_p$ and $x=\omega_1/(\omega_1+\omega_2)$.}
\label{G}
\end{figure}

In  Fig.~\ref{G} we show the dependence of the function $G[\Omega x,\Omega (1-x)]$ on $x$ at fixed values of $\Omega$, where $\Omega=(\omega_1+\omega_2)/\varepsilon_p$ and $x=\omega_1/(\omega_1+\omega_2)$.

Within logarithmic accuracy we also have for the Born cross section
\begin{align}\label{spectrumB}
&d\sigma^B=\frac{8\alpha^2\eta^2d\omega_1d\omega_2}{\pi m^2\omega_1\omega_2}\,G(\omega_1/\varepsilon_p,\, \omega_2/\varepsilon_p )\,\ln\frac{1}{\beta_0}\,,
\end{align}
where  the function $G$ is the same  as in Eq.~\eqref{spectrumCC},  and
\begin{equation}\label{beta0}
\beta_0=\mbox{max}\left\{\frac{1}{mr_{scr}}\,,\, \frac{ m(\omega_1+\omega_2)}{\varepsilon_p \varepsilon_q}\right\}\ll 1\,.
\end{equation}

\section{Conclusion}\label{concl}
We have investigated in detail  the  process  of  high-energy double  bremsstrahlung  in the  field of a heavy atom.  The results, Eq. \eqref{M12F}, are  exact in the parameters of the atomic field and are  valid even for $\eta\sim 1$. The Coulomb corrections to the differential cross section are very sensitive to the shape of the atomic potential,  while the Coulomb corrections  to the  cross section, integrated over the momentum transfer $\bm\Delta_\perp$, are the universal function of $\eta$.  It is shown that, similar to the case of single bremsstrahlung, the potential enters the amplitudes of  high-energy double  bremsstrahlung via the factor $\bm A(\bm\Delta_\perp)$. Note that such factorization takes place only for the cross section obtained in the leading quasiclassical approximation and is violated by the first quasiclassical correction. It follows from the result of Ref.~\cite{KM2015} that the main contribution to the first quasiclassical correction to the cross section is given by the region $\Delta\sim m$. The factorized form of the amplitudes \eqref{M12F} and also of the amplitudes of single bremsstrahlung allows us to formulate the recipe for the calculation of the multiple bremsstrahlung differential cross section. In order to obtain the amplitude of this process exactly in the parameter $\eta$ for any shape of the atomic potential $V(r)$, it is sufficient to derive the amplitude in the Born approximation and then to replace in this amplitude the  Fourier transform  $V_F(\Delta^2)$ of the potential $V(r)$ by the impact-factor $A_0(\bm\Delta_\perp)$ \eqref{AA0}. Our recipe extends the impact-factor approach of Ref. \cite{CW1970} to the region of small momentum transfer. Note that it is just the region where the Coulomb corrections to the cross section of bremsstrahlung come from. We stress that our formulas for the cross sections of  high-energy double bremsstrahlung are obtained exactly in the parameter $\eta=Z\alpha$ and, in particular, valid for $Z\gg 1$. This is important for analysis of experimental data from modern detectors, where high-$Z$ materials are widely used.

\section*{Acknowledgement}
This work has been  supported by Russian Science Foundation (Project N 14-50-00080).

\section*{Appendix}
In this Appendix, following the method of \cite{LMSS2005}, we consider the calculation of the quantity ${\cal M}$,
\begin{align}\label{calm}
{\cal M}&=\iint d\bm r_1d\bm r_2\,e^{-i\bm k_1\cdot\bm r_1-i\bm k_2\cdot\bm r_2 }f_0(\bm r_2)d_0(\bm r_2,\bm r_1)g_0(\bm r_1)\,,
\end{align}
which contributes to the amplitude \eqref{MF}. Other quantities  are calculated in the same way.
The functions $d_0 $, $f_0 $, and $g_0$ are  given in Eq.~\eqref{dqc}, Eq.~\eqref{fgqc}, and Eq.~\eqref{gvf}, respectively.

We split the integration region into three, $z_1<z_2<0$, $z_1<0\,\&\,z_2>0$, $z_2>z_1>0$, and denote the corresponding contributions to ${\cal M}$   as  ${\cal M}_1$,  ${\cal M}_2$, and
${\cal M}_3$. In the first region,  the functions $g_0$ and $d_0$ have  simple  eikonal forms
\begin{align}\label{dgeik}
&g_0(\bm r_1)=e^{i\bm p\cdot\bm r_1}\exp\left[-i\int_0^\infty dx V(\bm r_1-x\bm n_{\bm p}) \right]\,,\nonumber\\
&d_0(\bm r_2,\bm r_1)=-\frac{e^{i\kappa r} }{4\pi r}\exp\left[-ir\int_0^1 dx V(\bm r_1+x\bm r) \right]\,,\nonumber\\
&\bm r=\bm r_2-\bm r_1\,,\quad \kappa=\sqrt{(\varepsilon_p-\omega_1)^2-m^2}\,,
\end{align}
so that
\begin{align}\label{M1}
{\cal M}_1&=\frac{i}{(2\pi)^2}\iint\limits_{z_1<z_2<0}\frac{ d\bm r_1d\bm r_2}{r}\,\int \,d\bm Q\, \exp(i\Phi)\,, \nonumber\\
&\Phi=Q^2 +(\bm p-\bm k_1)\cdot\bm r_1-(\bm q+\bm k_2)\cdot\bm r_2+\kappa r \nonumber\\
&-\int_0^\infty dx\,V(\bm r_1-x\bm n_{p})-r\int_0^1 dx V(\bm r_1+x\bm r)-\int_0^\infty dx\,V(\bm r_q)\,,\nonumber \\
&\bm r_q=\bm r_2+x\bm n_q+\bm Q\sqrt{\frac{2|\bm n_q\cdot\bm r_2|}{q}}\,.
\end{align}
Within our accuracy we can  replace   the quantity $V(\bm r_1-x\bm n_{p})$  and  $V(\bm r_1+x\bm r)$  in \eqref{M1} by $V(\bm r_1-x\bm n_{p}+\bm Q\sqrt{2|\bm n_q\cdot\bm r_2|/q})$ and
$V(\bm r_1+x\bm r+\bm Q\sqrt{2|\bm n_q\cdot\bm r_2|/q}\,)$, respectively,   shift $\bm \rho_1\to\bm \rho_1-\bm Q\sqrt{2|\bm n_q\cdot\bm r_2|/q}$, $\bm \rho_2\to\bm \rho_2-\bm Q\sqrt{2|\bm n_q
\cdot\bm r_2|/q}$, where
$\bm\rho_1=\bm r_{1\perp}$ and  $\bm\rho_2=\bm r_{2\perp}$. Then we  take the integral over $\bm Q$ and  obtain
\begin{align}\label{M11}
{\cal M}_1&=-\frac{1}{4\pi}\iint\limits_{z_1<z_2<0}\frac{ d\bm r_1d\bm r_2}{r}\, \exp[i(\Phi_0+\Phi_1)]\,, \nonumber\\
&\Phi_0=(\bm p-\bm k_1)\cdot\bm r_1-(\bm q+\bm k_2)\cdot\bm r_2+\kappa r \nonumber\\
&-\int_0^\infty dx\,V(\bm r_1-x\bm n_{p})-r\int_0^1 dx V(\bm r_1+x\bm r)-\int_0^\infty dx\,V(\bm r_2+x\bm n_q)\,,\nonumber\\
&\Phi_1=-\frac{\Delta_\perp^2|\bm n_q\cdot\bm r_2|}{2q}\,.
\end{align}
In the same way, we obtain for ${\cal M}_2$ and  ${\cal M}_3$,
\begin{align}\label{M2}
{\cal M}_2&=-\frac{1}{4\pi}\iint\limits_{z_2>0,z_1<0}\frac{ d\bm r_1d\bm r_2}{r}\, \exp[i(\Phi_0+\Phi_2)]\,, \nonumber\\
&\Phi_2=-\frac{\Delta_\perp^2|\bm r\cdot\bm r_1||\bm r\cdot\bm r_2|}{2\kappa r^3}\,,\nonumber\\
{\cal M}_3&=-\frac{1}{4\pi}\iint\limits_{z_2>z_1>0}\frac{ d\bm r_1d\bm r_2}{r}\, \exp[i(\Phi_0+\Phi_3)]\,, \nonumber\\
&\Phi_3=-\frac{\Delta_\perp^2|\bm n_p\cdot\bm r_1|}{2p}\,.
\end{align}
There are two overlapping regions of the momentum transfer $\Delta$:
\begin{align}\label{regions}
&\mbox{I}.\, \Delta\gg\frac{m^2(\omega_1+\omega_2)}{\varepsilon_p\varepsilon_q}\nonumber\\
&\mbox{II}.\, \Delta\ll\frac{m(\omega_1+\omega_2)}{\varepsilon_p}\,.
\end{align}
In the first region we can neglect in the phase $\Phi_0$ the term $\Delta_\parallel$ as compared with $\Delta_\perp$ and replace in the integrals $\bm n_q\rightarrow\bm\nu$ and
 $\bm r\rightarrow (\bm \nu\cdot\bm r)\bm\nu$, where z axes is parallel to $\bm\nu=\bm n_p$.  Performing the integration over $z_1$, $z_2$, and $\bm\rho_2-\bm\rho_1$, we obtain
\begin{align}\label{calmItot}
&{\cal M}=\frac{i}{2}\int d\bm \rho\exp\left[-i\bm\Delta_\perp\cdot\bm\rho-i\chi(\rho)\right][qN_1-\kappa N_2-pN_3]\,,
\end{align}
where the quantities $\chi(\rho)$, $N_1$, $N_2$, and $N_3$ are defined in Eq.~\eqref{notation}. Then we use the relation
\begin{align}\label{rel}
&qN_1-\kappa N_2-pN_3=\bm\Delta_\perp\cdot\bm j_0\,,
\end{align}
where  $\bm j_0$ is given in Eq.~\eqref{notation}. Performing integration by parts, we finally obtain ${\cal M}$ in the first region:
\begin{align}\label{calmItotf}
&{\cal M}=-\frac{i}{2}\int d\bm \rho\exp\left[-i\bm\Delta_\perp\cdot\bm\rho-i\chi(\rho)\right]\bm\nabla_\perp\chi(\rho)\cdot\bm j_0 \nonumber\\
&=-\frac{i}{2}\int d\bm r\exp\left[-i\bm\Delta_\perp\cdot\bm\rho-i\chi(\rho)\right]\bm\nabla_\perp V(r)\cdot\bm j_0 \,.
\end{align}
In the second region, one can neglect the term $\Phi_1$ in Eq.~\eqref{M11} and the terms  $\Phi_{2,3}$ in Eq.~\eqref{M2}. In the phase $\Phi_0$  we  take into account the linear terms of  expansion of the integrals in  $\bm n_{q}-\bm\nu$ and  $\bm r-(\bm r\cdot\bm\nu)\bm\nu$. The result, which is valid both in the region I and in the region II, has the form
\begin{align}\label{calmItotff}
&{\cal M}=-\frac{i}{2}\int d\bm r\exp\left[-i\bm\Delta\cdot\bm r-i\chi(\rho)\right]\bm\nabla_\perp V(r)\cdot\bm j_0 \,.
\end{align}
It corresponds to the second line in Eq.~\eqref{calmItotf} with the replacement $\bm\Delta_\perp\cdot\bm\rho\rightarrow\bm\Delta\cdot\bm r$.

 \end{document}